\input harvmac
\input epsf
\noblackbox
%%%%%%%%%%%%%%%%%%%%%%%%%%%%%%%%%%%%%%%%%%%%%%%%%%%%%%%%%%%%%%%%%%%%%%%%%

%%%%%%%%%%%%%%%%%%%%%%%%%%%%%%%%%%%%%%%%%%%%%%%%%%%%%%%%%%%%%%%%%%%%%%%%%
% some stuff needed for figures:
%%%%%%%%%%%%%%%%%%%%%%%%%%%%%%%%%%%%%%%%%%%%%%%%%%%%%%%%%%%%%%%%%%%%%%%%%
\newcount\figno
\figno=0
\def\fig#1#2#3{
\par\begingroup\parindent=0pt\leftskip=1cm\rightskip=1cm\parindent=0pt
\baselineskip=11pt
\global\advance\figno by 1
\midinsert
\epsfxsize=#3
\centerline{\epsfbox{#2}}
\vskip 12pt
{\bf Fig.\ \the\figno: } #1\par
\endinsert\endgroup\par
}
\def\figlabel#1{\xdef#1{\the\figno}}
\def\encadremath#1{\vbox{\hrule\hbox{\vrule\kern8pt\vbox{\kern8pt
\hbox{$\displaystyle #1$}\kern8pt}
\kern8pt\vrule}\hrule}}
%%%%%%%%%%%%%%%%%%%%%%%%%%%%%%%%%%%%%%%%%%%%%%%%%%%%%%%%%%%%%%%%%%%%%%%%%

%%%%%%%%%%%%%%%%%%%%%%%%%%%%%%%%%%%%%%%%%%%%%%%%%%%%%%%%%%%%%%%%%%%%%%%%%
\def\bberkeley{\centerline{\it Berkeley Center for Theoretical Physics and 
Department of Physics}
\centerline{\it  University of California, Berkeley, CA 94720-7300}
\centerline{\it and}
\centerline{\it Theoretical Physics Group, Lawrence Berkeley National 
Laboratory}
\centerline{\it Berkeley, CA 94720-8162, USA}}
%%%%%%%%%%%%%%%%%%%%%%%%%%%%%%%%%%%%%%%%%%%%%%%%%%%%%%%%%%%%%%%%%%%%%%%%%
\font\authfont=cmr12

%%%%%%%%%%%%%%%%%%%%%%%%%%%%%%%%%%%%%%%%%%%%%%%%%%%%%%%%%%%%%%%%%%%%%%%%%
\def\tilde#1{\widetilde{#1}}
%%%%%%%%%%%%%%%%%%%%%%%%%%%%%%%%%%%%%%%%%%%%%%%%%%%%%%%%%%%%%%%%%%%%%%%%%
\def\sign{\,{\rm sign}\,}
\def\frac#1#2{{#1 \over #2}}

\def\p{\partial}

\def\CD{{\cal D}}

\def\CM{{\cal M}}                   \def\CN{{\cal N}}
\def\CO{{\cal O}}

\def\R{{\bf R}}                     
                     
%

%%%%%%%%%%%%%%%%%%%%%%%%%%%%%%%%%%%%%%%%%%%%%%%%%%%%%%%%%%%%%%%%%%%%%%%%%%%
%%%%%%%%%%%%%%%%%%%%%%%%%%%%%%%%%%%%%%%%%%%%%%%%%%%%%%%%%%%%%%%%%%%%%%%%%%%
\Title{\vbox{\baselineskip12pt
\hbox{hep-th/0502127}
\hbox{\ }}}
{\centerline{Topology Changing Transitions in Bubbling Geometries}}
%%%%%%%%%%%%%%%%%%%%%%%%%%%%%%%%%%%%%%%%%%%%%%%%%%%%%%%%%%%%%%%%%%%%%%%%%%%
\bigskip
\centerline{\authfont Petr Ho\v rava and Peter G. Shepard}
\medskip\bigskip\medskip
\baselineskip14pt
\bberkeley
\centerline{\tt horava, pgs@socrates.berkeley.edu}
\medskip\bigskip\medskip\bigskip
\centerline{\bf Abstract}
\bigskip
\noindent
Topological transitions in bubbling half-BPS Type IIB geometries with 
$SO(4)\times SO(4)$ symmetry can be decomposed into a sequence of $n$ 
elementary transitions.  The half-BPS solution that describes 
the elementary transition is seeded by a phase space distribution of fermions 
filling two diagonal quadrants.  We study the geometry of this solution in 
some detail.  We show that this solution can be interpreted as a time 
dependent geometry, interpolating between two asymptotic pp-waves in the 
far past and the far future.  The singular solution at the transition can 
be resolved in two different ways, related by the particle-hole duality in 
the effective fermion description.  Some universal features of the topology 
change are governed by two-dimensional Type 0B string theory, whose double 
scaling limit corresponds to the Penrose limit of $AdS_5\times S^5$ at 
topological transition.  In addition, we present the full class of 
geometries describing the vicinity of the most general localized classical 
singularity that can occur in this class of half-BPS bubbling geometries.  
\Date{February 2005}
%%%%%%%%%%%%%%%%%%%%%%%%%%%%%%%%%%%%%%%%%%%%%%%%%%%%%%%%%%%%%%%%%%%%%%%%%%%
\nref\llm{H. Lin, O. Lunin and J. Maldacena, ``Bubbling AdS Space and 1/2 BPS 
Geometries,'' hep-th/0409174.}
\nref\antal{S. Corley, A. Jevicki and S. Ramgoolam, ``Exact Correlators of 
Giant Gravitons from Dual $N=4$ SYM,'' hep-th/0111222.}
\nref\db{D. Berenstein, ``A Toy Model of the AdS/CFT Correspondence,'' 
hep-th/0403110.}
\nref\akietal{A. Hashimoto, S. Hirano and N. Itzhaki, ``Large Branes in 
AdS and Their Field Theory Dual,'' hep-th/0008016.}
\nref\calda{M.M. Caldarelli and P.J. Silva, ``Giant Gravitons in AdS/CFT (I): 
Matrix Model and Back Reaction,'' hep-th/0406096.}
\nref\harmstr{N. Itzhaki and J. McGreevy, ``The Large $N$ Harmonic Oscillator 
as a String Theory,'' hep-th/0408180\hfill\break
A. Boyarsky, V.V. Cheianov and O. Ruchayskiy, ``Fermions in the Harmonic 
Potential and String Theory,'' hep-th/0409129.}
\nref\crev{I. Klebanov, ``String Theory in Two Dimensions,'' hep-th/9108019
\hfill\break
G. Moore and P. Ginsparg, ``Lectures on 2D Gravity and 2D String Theory,'' 
hep-th/9304011.}
\nref\chat{M.R. Douglas, I.R. Klebanov, D. Kutasov, J. Maldacena, E. Martinec 
and N. Seiberg, ``A New Hat for the $c=1$ Matrix Model,'' hep-th/0307195
\hfill\break
T. Takayanagi and N. Toumbas, ``A Matrix Model Dual of Type 0B String Theory 
in Two Dimensions,'' hep-th/0307083.}
\nref\vafaetc{D. Ghoshal and C. Vafa, ``$c=1$ String as the Topological Theory 
on the Conifold,'' hep-th/9506122.\hfill\break
H. Ooguri and C. Vafa, ``Two-Dimensional Black Hole and Singularities of 
CY Manifolds,'' hep-th/9511164.}
\nref\givkut{A. Giveon and D. Kutasov, ``Little String Theory in a Double 
Scaling Limit,'' hep-th/9909110; ``Comments on Double Scaled Little String 
Theory,'' hep-th/9911039.}
\nref\rhdcv{R. Dijkgraaf and C. Vafa, ``$N=1$ Supersymmetry, Deconstruction, 
and Bosonic Gauge Theories,'' hep-th/0302011.}
\nref\minaetal{M. Aganagic, R. Dijkgraaf, A. Klemm, M. Mari\~no and C. Vafa, 
``Topological Strings and Integrable Hierarchies,'' hep-th/0312085.}
\nref\frag{J. Maldacena, J. Michelson and A. Strominger, ``Anti-de~Sitter 
Fragmentation,'' hep-th/9812073.}
\nref\gubser{S.S. Gubser, ``Curvature Singularities: The Good, the Bad, and 
the Naked,'' hep-th/0002160.}
%%%%%%%%%%%%%%%%%%%%%%%%%%%%%%%%%%%%%%%%%%%%%%%%%%%%%%%%%%%%%%%%%%%%%%%%%%%
\newsec{Introduction}

An intriguing picture of half-BPS geometries of Type IIB string theory 
corresponding to the chiral primaries (with $\Delta=J$) of $\CN=4$ SYM theory 
has emerged recently in the work of Lin, Lunin and Maldacena (LLM) \llm .  
In this picture, the data defining the geometry is captured by a distribution 
of incompressible droplets of fermions on a $1+1$ dimensional phase space.  
Since the effective $\hbar$ of the fermions is related to the Planck length 
of Type IIB supergravity via $\hbar=2\pi\ell_p^4$, the semiclassical limit 
of the Fermi system -- in which it makes sense to talk about the geometry of 
the Fermi surface in phase space -- corresponds to the semiclassical limit of 
the Type IIB geometry.  On the CFT side, the fermionic picture of the 
geometries nicely matches the fermions emerging as the eigenvalues in 
the gauged matrix model description of the corresponding half-BPS states 
with $\Delta=J$ (schematically of the form $\Pi_i\Tr(Z^{n_i})^{r_i}$), 
in the harmonic oscillator potential \refs{\antal-\calda} (see also 
\harmstr ).  

Generically, an individual classical geometry corresponds to some distribution 
of fermions with a number of boundaries between occupied 
and unoccupied regions.  Zooming in on a small part of the boundary between 
the two regions corresponds to a pp-wave limit of the geometry, and thus 
the geometry is nonsingular in the vicinity of such boundary.  

As we move around in the moduli space of such geometries, the phase space 
liquid will flow while preserving its total area.  In the process, the number 
of components of the Fermi surface separating the filled and empty regions 
can change.  In the corresponding Type IIB geometry, this represents a 
spacetime topology change.  

Generically, this change will happen by two boundaries approaching each other 
and reconnecting with a net change of boundary components by $\pm 1$ (see 
Fig.~1).  
\fig{Two semiclassical phase space distributions of fermions, related by a 
topological transition that changes the number of components of the Fermi 
surface by $\pm 1$.}{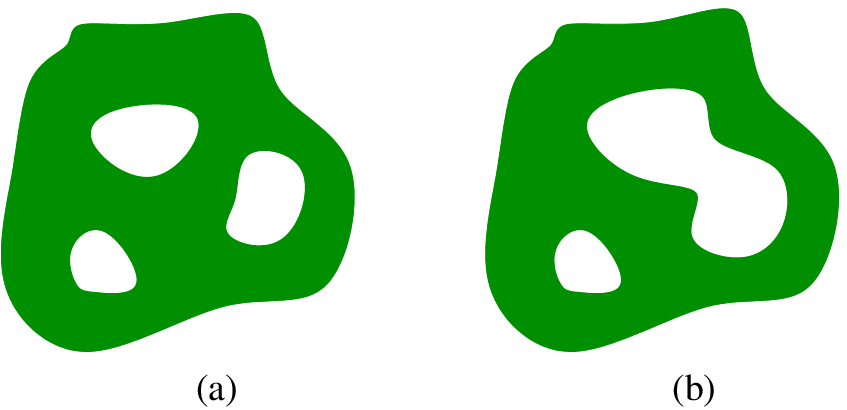}{3truein}

More complicated topology changing processes in bubbling geometries are also 
possible: for example, one can envision more than two boudary components 
meeting and reconnecting.  However, since the phase space of the fermions 
is only two-dimensional, any such process is non-generic and can be smoothly 
deformed into a sequence of steps, each step only involving the elementary 
process of two boundaries reconnecting.%
\foot{A full class of non-generic localized topology changing processes is 
discussed in Section~6.}

We wish to understand the geometry at the transition when two boundary 
components meet and reconnect.  The universal features of the transition 
will be captured by the local region where the two boundary components touch.  
We will isolate these universal features by zooming in on the intersection.  
The fermions that source this geometry occupy two opposite infinite quadrants 
on phase space, as in Fig.~2.%
\foot{ For simplicity, we shall concentrate on the case of Fermi surface 
components meeting at right angles throughout most of the paper, relegating 
the more general case of arbitrary angles to Section~6.}
All other topology changing processes in the bubbling Type IIB geometries of 
LLM are then decomposable into $n$ steps, each involving the unique topology 
changing process studied here.  
\fig{The phase space distribution of fermions that seeds the universal part of 
the geometry at the topological transition.}{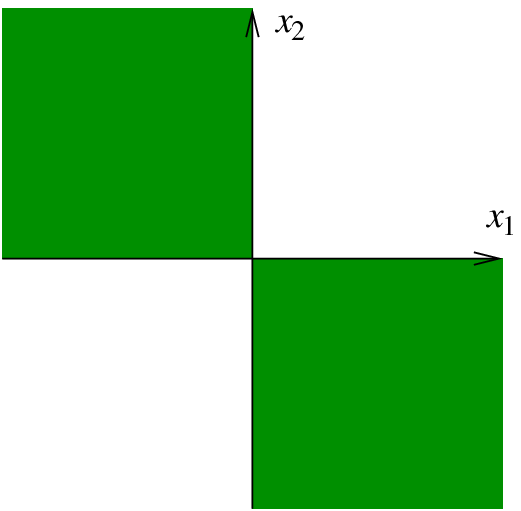}{2truein}

This paper is organized as follows.  In Section~2, we present the univesal 
topology changing solution and study some of its properties.  In Section~3 we 
reinterpret this solution as a time-dependent process.  In Section~4 we 
study resolutions to non-singular geometries with the same asymptotic 
behavior; here the non-critical $\hat c=1$ Type 0B strings make 
an appearance.  In Section~5 we embed the topology changing process into 
$AdS_5\times S^5$, refining our understanding of how the universal properties 
of the topology change are captured by the double scaling limit known to 
define the noncritical Type 0B string theory in two dimensions.  In Section~6 
we consider the complete class of half-BPS LLM geometries of Type IIB theory 
with the additional property of scaling; these geometries describe the most 
general composite classical singularities that can occur in the 
``quantum-foam'' topological transitions in LLM geometries. 

\newsec{The Half-BPS Solution at Topological Transition}

\subsec{LLM geometries}

It has been shown in \llm\ that the most general half-BPS solution of Type IIB 
supergravity with $SO(4)\times SO(4)\times\R$ symmetry, with only the metric 
and the five-form RR flux excited, can be written as 
\eqn\eesolmet{ds^2=-h^{-2}(dt+V_idx^i)^2+h^2(dy^2+dx^idx^i)+R^2d\Omega_3^2
+\tilde R^2d\tilde\Omega_3^2}
and
\eqn\eesolflux{F_5\equiv F_{\mu\nu}dx^\mu\wedge dx^\nu\wedge 
d\Omega_3+\tilde F_{\mu\nu}dx^\mu\wedge dx^\nu\wedge d\tilde\Omega_3,}
with the two two-forms $F_{\mu\nu}$ and $\tilde F_{\mu\nu}$ given by
\eqn\eesolfluxtwo{\eqalign{F&=-\frac{1}{4}\left\{d\left(R^4(dt+V)\right)+
y^3\ast_3d\left(\frac{z+\frac{1}{2}}{y^2}\right)\right\},\cr
\tilde F&=-\frac{1}{4}\left\{d\left(\tilde R^4(dt+V)\right)+y^3\ast_3d\left(
\frac{z-\frac{1}{2}}{y^2}\right)\right\}.\cr}}
Here $R$ and $\tilde R$ are the radii of the two three-spheres on which the 
two $SO(4)$'s act transitively (following \llm , we shall denote them simply 
by $S^3$ and $\tilde S^3$), and $t$ corresponds to the time coordinate with 
$\p/\p t$ the (mostly timelike) Killing vector that generates the $\R$ part of 
the symmetry group.  In Eqn.~\eesolflux\ for the fluxes, $x^\mu$ with $\mu=0,
\ldots 3$ refers collectively to the four coordinates $t,x_i,y$, and $\ast_3$ 
is the Hodge operator on the $x_i,y$ space with respect to its flat metric.  

All the coefficients $R,\tilde R, V_1,V_2$ and $h$ in \eesolmet\ are 
determined in terms of a single function $z$ on the three-dimensional space 
$x_1,x_2,y$. The radii $R$ and $\tilde R$ of the $S^3$ and $\tilde S^3$ 
factors of the geometry are given in terms of $z$ by
\eqn\eesolradii{R^2=y\sqrt{\frac{1+2z}{1-2z}},\qquad 
\tilde R^2=y\sqrt{\frac{1-2z}{1+2z}},}
the coefficient $h$ is given by
\eqn\eesolh{h^{-2}=\frac{2y}{\sqrt{1-4z^2}}=R^2+\tilde R^2,}
and the components of $V$ can be determined, once $z$ is known, from
\eqn\eesolvect{y\p_yV_i=\epsilon_{ij}\p_jz,\qquad
y(\p_iV_j-\p_jV_i)=\epsilon_{ij}\p_yz.}

The equations of motion of Type IIB supergravity are then equivalent 
to the following linear equation for the function $z$, 
\eqn\eezeeeq{\p_i\p_iz+y\p_y\left(\frac{\p_yz}{y}\right)=0,}
where $\p_i\p_i$ is the flat-space Laplacian on the $(x_1,x_2)$ plane.  
Since $y=R\tilde R$, $y$ takes only positive values, and \eezeeeq\ can 
be solved by first selecting an initial condition $z(x_1,x_2,0)$ and then 
solving for the $y$ dependence of $z$ by \eezeeeq .  The initial condition 
on $z$ at $y=0$ can be a priori arbitrary, but unless $z(y=0)$ is only equal 
to $\pm1/2$ the geometry is singular.  Thus, all solutions of the LLM type 
are classified by specifying the areas on the $x_1,x_2$ plane where $z$ is 
$1/2$ or $-1/2$; by reinterpreting these two regions as being empty or filled 
by semiclassical fermions, one can establish the correspondence between such 
half-BPS geometries and the corresponding half-BPS operators in the CFT dual 
(reduced to the matrix model).  

As an example, $AdS_5\times S^5$ itself corresponds to the circular 
Fermi surface.  Its Penrose limit corresponds to the Fermi sea filling one 
half of the phase space, with a linear Fermi surface; this is the Fermi 
description of the Type IIB pp-wave.  In both of these special 
examples, the Fermi sea develops an ``accidental'' extra symmetry compatible 
with the flat metric on the $(x_1,x_2)$ plane: rotations around the origin 
in the case of $AdS_5\times S^5$, or translations along the Fermi surface 
for the pp-wave.  In the full geometry, these extra symmetries give rise to 
additional Killing vectors.  

Conversely, one can interpret individual Killing vectors of the full Type IIB 
geometry in terms of flows on the phase space of the Fermi system: for 
example, the generator $H\equiv\Delta$ of time translations in the 
conventional global coordinates on $AdS_5\times S_5$ corresponds to the 
Hamiltonian flow that uniformly rotates the phase space around the center of 
the Fermi sea.  All geometries have at least one Hamiltonian flow associated 
with the Killing vector mandated by the presence of supersymmetry; in the 
terminology of the CFT dual, this Hamiltonian flow is generated by 
$H'=\Delta-J$, and acts trivially on the $x_1,x_2$ plane.  

\subsec{The solution at topological transition in LLM coordinates}

After this brief summary of the results of \llm , we can introduce the 
solution that captures the universal features of the unique generic 
topology-changing process in all half-BPS Type IIB geometries of this class. 
As discussed in the Introduction, this leads us to solve for $z$ sourced by 
fermions distributed in two opposite infinite quadrants (cf.\ Fig.~2),
\eqn\eesource{z(x_1,x_2,0)=\frac{1}{2}\sign(x_1)\sign(x_2).}
The solution of \eezeeeq\ with such initial conditions is 
\eqn\eesolz{z(x_1,x_2,y)=\frac{x_1}{\pi\sqrt{x_1^2+y^2}}
\arctan\left(\frac{x_2}{\sqrt{x_1^2+y^2}}\right)
+\frac{x_2}{\pi\sqrt{x_2^2+y^2}}\arctan\left(\frac{x_1}{\sqrt{x_2^2+y^2}}
\right).}
Given this $z$, one can also find the components of the vector field $V$, 
\eqn\eesolv{\eqalign{V_1(x_1,x_2,y)&=-\frac{1}{\pi\sqrt{x_2^2+y^2}}
\arctan\left(\frac{x_1}{\sqrt{x_2^2+y^2}}\right),\cr
V_2(x_1,x_2,y)&=\frac{1}{\pi\sqrt{x_1^2+y^2}}\arctan\left(
\frac{x_2}{\sqrt{x_1^2+y^2}}\right).\cr}}
Having determined $V$, one can -- in retrospect -- write $z$ in the following 
simple form, 
\eqn\eezviav{z(x_1,x_2,y)=x_1V_2-x_2V_1.}
The rest of the geometry, including the RR five-form fluxes, is then known 
in terms of $z$ as summarized in the previous subsection.  

Since one of the requirements in the construction of all LLM solutions is the 
existence of a sixteen-component generalized Killing spinor, our solution is 
guaranteed to have (at least) sixteen supersymmetries.  

\subsec{Scaling properties}

Since the phase space distribution of the fermions that seeds this geometry 
is scale invariant, the metric will inherit an interesting scaling property.  
Consider the following transformation, 
\eqn\eescaling{x_i\rightarrow\lambda x_i,\qquad y\rightarrow\lambda y.}
Under this tranformation, we have
\eqn\eefscaling{\eqalign{
\eqalign{z&\rightarrow z,\cr R&\rightarrow \lambda^{1/2}R,\cr}
&\qquad\eqalign{V_i&\rightarrow\lambda^{-1}V_i,\cr
\tilde R&\rightarrow \lambda^{1/2}\tilde R,\cr}\cr
h\rightarrow&\lambda^{-1/2}h.\cr}}
In particular, $z$ is invariant under this scaling transformation.  

If we now extend the action of our scaling transformation to $t$ and the six 
angle coordinates on $S^3$ and $\tilde S^3$ so that they are all scaling 
invariant, 
\eqn\eetscaling{t\rightarrow t,\qquad d\Omega_3^2\rightarrow d\Omega_3^2,
\qquad d\tilde\Omega_3^2\rightarrow d\tilde\Omega_3^2,}
the metric and the RR flux will transform as follows,
\eqn\eemetscaling{ds^2\rightarrow\lambda ds^2,\qquad 
F_5\rightarrow\lambda^2F_5.}
In Section~6 we will study a complete class of solutions which share the 
same scaling property.  

\subsec{Behavior near $y=0$}

Consider the nine-dimensional slice $\CM_9$ of constant $t$ in our solution 
at topological transition, as given by \eesolz , \eesolv , \eesolmet\ and 
\eesolflux .  Away from the $y=0$ hypersurface, the product of the two 
three-spheres $S^3\times\tilde S^3$ is trivially fibered over the 
topologically trivial open domain $(x_1,x_2,y)$, $y>0$.  In order to 
understand various properties of $\CM_9$, such as its global topology, 
we first need to determine the behavior at $y=0$.  The radii of the two 
spheres are given by $y\sqrt{\frac{1\pm 2z}{1\mp 2z}}$, Eqn.~\eesolradii .  
When we approach $y=0$, $z$ approaches $\pm1/2$.  In this limit, one of the 
radii will always go to zero, while the other one will generically stay 
finite.  In the case of our solution, the radius of the first $S^3$ is 
non-zero in the two quadrants where $\sign(x_1/x_2)=1$, and its value there is
\eqn\eeradyzero{R^2(y=0)=(2\pi|x_1x_2|)^{1/2}\left(1+\frac{x_1}{x_2}\arctan
\left(\frac{x_1}{x_2}\right)+\frac{x_2}{x_1}\arctan\left(\frac{x_2}{x_1}\right)
\right)^{-1/2}.}
Similarly, the radius of $\tilde S^3$ is also given by \eeradyzero , but now 
in the complementary quadrants satisfying $\sign(x_1/x_2)=-1$.  At the 
boundary between the two regions, both spheres shrink to zero volume, and 
the geometry is smooth everywhere outside of $x_i=y=0$.  

Having calculated the radii of $S^3$ and $\tilde S^3$ at $y=0$, we can now 
examine the norm of the ``mandatory'' Killing vector 
$\p/\p t$ whose existence follows from the BPS property of the solution,   
\eqn\eenorm{\eqalign{\left\|\frac{\p}{\p t}\right\|^2&=-R^2-\tilde R^2\cr
&=-(2\pi|x_1x_2|)^{1/2}\left(1+\frac{x_1}{x_2}\arctan\left(\frac{x_1}{x_2}
\right)+\frac{x_2}{x_1}\arctan\left(\frac{x_2}{x_1}\right)\right)^{-1/2}.\cr}}
This formula is valid everywhere on the $(x_1,x_2)$ plane at $y=0$.  Thus, 
the Killing vector $\p/\p t$ is time-like inside the $z=\pm 1/2$ regions, and 
turns null at the boundaries between the regions.  In particular, the 
singularity at the origin is a null singularity.  Since the radii of both 
$S^3$ and $\tilde S^3$ shrink to zero there, the singularity corresponds in 
the full ten-dimensional geometry to a null worldline of a point-like massless 
particle.  Outside of the $y=0$ hypersurface, both $R$ and $\tilde R$ are 
non-zero, and $\p/\p t$ is strictly timelike.  

\subsec{The topology of the solution}

In order to determine the global topology of $\CM_9$, we shall now study 
eight-dimensional closed surfaces inside $\CM_9$, constructed by fibering 
$S^3\times\tilde S^3$ over a two-dimensional disk $\Sigma_D$ embedded into the 
$x_1,x_2,y$ space such that $\p\Sigma_D$ is inside the $y=0$ surface, while 
the interior of $\Sigma_D$ is mapped into the region with $y>0$.  

There are three interesting cases to consider.  They correspond to $
\Sigma_D=\Sigma_1, \Sigma_2$ and $\Sigma_3$ in Fig.~3.  
\fig{The three eight-surfaces discussed in Section~2.5.  The surfaces 
obtained from the disks $\Sigma_1$, $\Sigma_2$ and $\Sigma_3$ are 
topologically $S^5\times S^3$, $S^8$ and $S^4\times S^4$, 
respectively.}{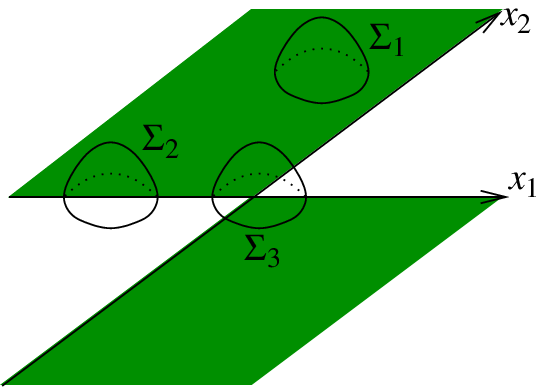}{3truein}

If $\p\Sigma_D$ lies entirely within the filled or empty region (as is the 
case for $\Sigma_D=\Sigma_1$ of Fig.~3), one of the 
three-spheres shrinks to zero volume uniformly along $\p\Sigma_D$ while the 
other $S^3$ stays at non-zero volume -- the resulting eight-surface is 
topologically $S^5\times S^3$ or $S^3\times S^5$.  

If $\p\Sigma_D$ intersects a smooth boundary component between the filled and 
empty region in two points (cf.~$\Sigma_2$ in Fig.~3), the resulting surface 
has the topology of $S^8$.  
This can be seen for example by first foliating $\Sigma_D$ by one-dimensional 
compact intervals $I$ such that one end-point of $I$ lies on $\p\Sigma_D$ 
inside the filled region while the other end-point lies on $\p\Sigma_D$ 
inside the empty region.  Over $I$, one $S^3$ shrinks to zero 
volume at one end-point of $I$, while the other $S^3 $ shrinks to zero volume 
at the other end-point of $I$.  Hence, $I$ together with the two three-spheres 
form an $S^7$. The total eight-surface costructred from $S^3$ and $\tilde S^3$ 
over $\Sigma_D$ is then a foliation of $S^8$ by $S^7$'s whose volume shrinks 
to zero at the two poles of $S^8$ where the surface intersects the boundary 
between the filled and empty region on the $(x_1,x_2)$ plane at $y=0$.  

Finally, when $\p\Sigma_D$ encircles the singular point at the origin, hence 
intersecting the boundary between the filled and empty regions in four points, 
the resulting surface is $S^4\times S^4$.  This can be again demonstrated by 
an argument similar to the one used in the previous paragraph.  The boundary 
$\p\Sigma_D$ of the disk is now divided into four segments; inside each 
segment, one $S^3$ has non-zero volume while the other one has shrunk to 
a point.  The two $S^3$'s exchange their roles as we move from segment to 
segment along $\p\Sigma_D$.  Imagine cutting the disk $\Sigma_D$ along a 
line connecting two boundary points on the two opposite segments of the 
boudary where the same $S^3$ has non-zero volume.  This cut splits the 
eight-manifold that results from fibefing the two $S^3$ over $\Sigma_D$ into 
two open eight-manifolds.  These open eight-manifolds are each topologically 
$D_4\times S^4$, and they are glued together along their $S^3\times S^4$ 
boundary so as to form $S^4\times S^4$.  The two noncontractible $S^4$ cycles 
on this surface are given by the four-manifolds obtained from considering one 
of the boundary segments on $\p\Sigma_D$ and fibering the $S^3$ of non-zero 
volume over it.  

The scaling property of our metric then implies that the full topology of 
the surfaces $\CM_9$ of constant $t$ in our solution is that of a cone over 
$S^4\times S^4$.  While the metric outside of the tip of the cone is smooth, 
there is a singularity at the tip of the cone at $x_i=y=0$.  

\subsec{Asymptotics to pp-waves}

There are regimes where our geometry approaches that of a pp-wave.  Indeed, 
consider the metric in the limit where (say) $x_1\gg |x_2|,|y|$.  
In that limit, we obtain
\eqn\eezeepplimit{z(x_1,x_2,y)\approx \frac{x_2}{2\sqrt{x_2^2+y^2}}
\quad+\CO\left(x_2/x_1,y/x_1\right),}
and
\eqn\eeveepplimit{\eqalign{V_1&\approx -\frac{1}{2\sqrt{x_2^2+y^2}}
\quad+\CO\left(\sqrt{x_2^2+y^2}/x_1\right),\cr
V_2&\approx \quad 0+\CO\left(x_2/x_1,y/x_1\right),\cr}}
(and similarly for the rest of the geometry including the RR fluxes.)
This is the maximally symmetric pp-wave solution of Type IIB 
theory, rewritten in the LLM coordinates.  The standard form of the 
pp-wave geometry is obtained by introducing coordinates 
\eqn\eeppcoords{r_1=R=\sqrt{x_2+\sqrt{x_2^2+y^2}},\qquad
r_2=\tilde R=\sqrt{-x_2+\sqrt{x_2^2+y^2}},}
and $x_1$.  In these coordinates, the metric becomes
\eqn\eemetppplus{\eqalign{ds^2=-2dtdx_1-(r_1^2+r_2^2)dt^2+dr_1^2+r_1^2
d\Omega_3^2+&dr_2^2+r_2^2d\tilde\Omega_3^2\cr
&{}+\CO(r_1^2/x_1,r_2^2/x_1).\cr}}
Notice that in this asymptotic pp-wave regime, $\p/\p x_1$ becomes an 
asymptotic Killing vector.   

Similarly, we could have considered any of the four regimes $\pm x_1\gg |x_2|, 
|y|$ or $\pm x_2\gg|x_1|,|y|$; in each of these regimes, our geometry is 
asymptotic to a pp-wave.  As a result of this asymptotic behavior, our 
solution develops extra asymptotic Killing symmetries and supersymmetries 
in these asymptotic regions. 

\newsec{The Geometry at Topological Transition as a Time-Dependent Process}

Until now we studied the solution in the LLM coordinates, in which it is 
manifestly stationary.  These coordinates are natural from the point of view 
of observers following the orbits of the (mostly) timelike Killing vector 
$K\equiv \p/\p t$.  This Killing vector indeed plays a crucial role in the 
LLM geometries, since it corresponds to $H'=\Delta-J$ on the CFT side, and 
supercharges square to $K$.  

We will now show that in our topology-changing geometry, a different class 
of natural observers can be introduced, for which the geometry asymptotes to 
a more conventional pp-wave geometry in the far past and the far future, 
and appears time-dependent at intermediate times.  

The fact that different observers can perceive a given geometry differently 
is indeed one of the crucial features of quantum gravity.  Making sense of the 
observations of the same geometry by different observers is one of the most 
intriguing challenges in cases such as de~Sitter space or the Schwarzschild 
black hole.  Our hope is that addressing such questions first in the more 
controlled case of supersymmetric geometries (such as those studied in this 
paper) might lead to important lessons with a broader range of applicability.  

\subsec{Relation to pp-waves}

As we showed in Section~2.6, our solution is asymptotic to the maximally 
supersymmetric Type IIB pp-wave, in the four regimes where $|x_1|$ or $|x_2|$ 
is much larger than the other coordinates on the $x_i,y$ space.  
We now introduce a new coordinate system, replacing $x^\mu$ with 
\eqn\eetdepcoor{\eqalign{t&=t',\cr
y&=r_1r_2,\cr
x_2&=\frac{1}{2}(r_1^2-r_2^2),\cr
x_1&=w+t'.\cr}}
These coordinates have been designed such that, asymptotically in the far 
past $t'\rightarrow-\infty$, our solution is asymptotic to the maximally 
supersymmetric pp-wave.  Thus, at early times in this coordinate system, we 
have
\eqn\eeastimemin{\eqalign{ds^2\approx-2dt'dw-[2+r_1^2+r_2^2]dt'^2+dr_1^2
+r_1^2d\Omega_3^2+&dr_2^2+r_2^2d\tilde\Omega_3^2\cr&{}
+\CO(r_1^2/t',r_2^2/t',w/t').\cr}}
The (noninertial) observers who are static with respect to this coordinate 
system will follow the lines of 
\eqn\eetimevec{\frac{\p}{\p t'}=\frac{\p}{\p t}+\frac{\p}{\p x_1}.}
In the asymptotic regime $t'\rightarrow -\infty$, becomes a Killing vector.  
However, since $\p/\p x_1$ is a Killing vector only asymptotically as 
$x_1\rightarrow-\infty$, $\p/\p t'$ will not be a Killing vector at finite 
$t'$: the observers will see a time-dependent geometry.  In the far past, 
such observers start in one of the asymptotic, asymptotically stationary 
pp-wave regimes, $x_1\rightarrow-\infty$.  

We can similarly define observers who see a geometry that asymptotes to 
a pp-wave in their far future.  Introduce coordinates similar to those in 
\eetdepcoor , 
\eqn\eetdepcoortwo{\eqalign{t&=t',\cr
y&=r_1r_2,\cr
x_1&=\frac{1}{2}(r_1^2-r_2^2),\cr
x_2&=w-t'.\cr}}
The observers who follow $\p/\p t'$ in these coordinates will end up in the 
pp-wave asymptotic regime $x_2\rightarrow\infty$ at late times, 
$t'\rightarrow\infty$.  

The observers introduced in this subsection see our topology-changing half-BPS 
solution quite differently from the stationary observers who follow the 
worldlines of the Killing vector $K$.  From their perspective, the geometry 
is more naturally interpreted as a time-dependent, background-scattering 
process, with a well-defined in- and out- asymptotics given by the maximally 
supersymmetric pp-wave.  It would be natural for such observers to probe the 
geometry by weakly interacting probes.%
\foot{In particular, if such probes are to maintain the half-BPS property of 
the geometry, they will be mutually non-interacting.}
The natural in-states and out-states are the excitations of the pp-wave 
geometry, and the natural observables are S-matrix-like elements (or 
reflection/transmission coefficients) for the scattering of such quanta off 
of  the time-dependent geometry.  The high degree of solvability of string 
theory in the asymptotic regimes, and the fact that our solution is half-BPS, 
suggest that a much more detailed analysis should be possible.  This is, 
however, outside of the scope of this paper.  

\subsec{Matching the asymptotic pp-wave regimes}

The time-dependent interpretation of our solution takes advantage of the 
pp-wave asymptotics, but it does so only in one pp-wave regime at a time.  
For example, the coordinate system \eetdepcoor\ is really natural at 
early times, but becomes less well-behaved at late times.  A different 
coordinate system, capable of capturing all four asymptotic pp-wave regimes, 
would be desirable.  

In order to motivate such a coordinate system, note first that the Fermi 
surface that seeds our geometry satisfies a simple equation $x_1x_2=0$.  
This can be considered a limit as $\mu\rightarrow 0$ of a smooth, hyperbolic 
Fermi surface $x_1x_2=\mu$, a fact that will take further advantage of in the 
next section. This entire family of Fermi surfaces enjoys an extra symmetry, 
suggesting the possibility of following a hyperbolic-type flow.  Such a flow 
will interpolate between the far-past and the far-future pp-wave asymptotics.  

The most straigthforward hyperbolic flow -- following the lines of constant 
$x_1x_2$ -- would not be a suitable choice, however.  This flow does 
interpolate bewteen the regions of $x_1\gg x_2$ and $x_2\gg x_1$, but it 
does not produce a regular coordinate system in the asymptotic limit.  
However, there is a natural ``regularization'' of the hyperbolic flow, 
which is in fact encoded in the geometry of our solution.  Consider the 
one-form $V=V_idx^i$ that appears in the metric.  The corresponding dual 
vector $\hat V$ is given by
\eqn\eevector{\hat V=h^{-2}\left(V_i\frac{\p}{\p x_i}-(V_iV_i)\frac{\p}{\p t}
\right).}
We will denote by $\tilde V$ the projection of $\hat V$ onto the surfaces of 
constant $t$.  The flow of $\tilde V$ on the $x_1,x_2$ plane (cf.\ Fig.~4) 
has exactly the properties we look for: 
\fig{The flow of $\tilde V$ on the $(x_1,x_2)$ plane at $y=0$ in the LLM 
coordinates.}{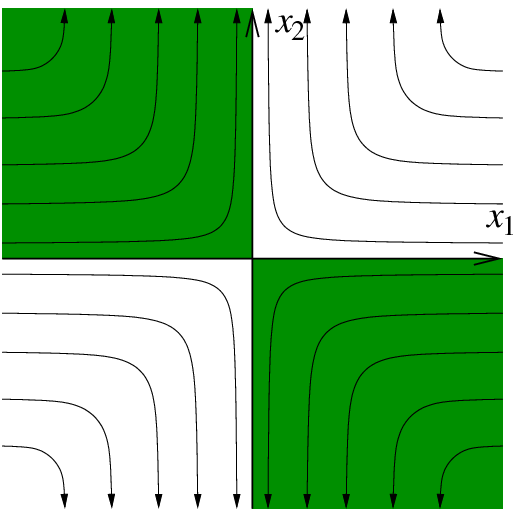}{2.3truein}

\item{(1)} The flow of $\tilde V$ interpolates between the asymptotic regions 
with large $|x_1|$ in the far past and the asymptotic regions with large 
$|x_2|$ in the far future,

\item{(2)} $\tilde V$ approaches the asymptotic Killing vectors in the 
corresponding asymptotic regions; for example, 
\eqn\eehatapp{\tilde V\approx-\frac{\p}{\p x_1}\qquad{\rm for}
\ x_1\rightarrow\infty.}

These observations suggest a particularly natural coordinate system to 
interpolate between the time-dependent interpretations of the solution in the 
four asymptotic pp-wave regimes.  Instead of following the lines of constant 
$x_1$ or $x_2$, the observers follow the regulated hyperbolic flow lines of 
$\tilde V$.  In such coordinates, the geometry is time dependent, and 
interpolates between two widely separated pp-waves (regimes of 
$|x_1|\rightarrow\infty$) in the asymptotic past, and two widely separated 
pp-waves (regimes with $|x_2|\rightarrow\infty$) in the asymptotic future.  
This picture again suggests that a natural interpretation of our solution 
is in terms of a scattering process.  

\newsec{Resolutions}

Until now we studied the half-BPS solution directly at the point of the 
topological transition.  This solution has a null point-like singularity at 
the origin, and the hypersurfaces of constant $t$ are topologically isomorphic 
copies of a cone over $S^4\times S^4$.  In this section, we will study the 
smooth resolutions of this singularity, within the family of half-BPS LLM 
geometries.   In the language of the fermions, such resolutions are simply 
accomplished by separating the Fermi surface into two smooth, disconnected 
components (cf.\ Fig.~5).  This can be done in two ways, leading to a 
one-parameter family of solutions interpolating between the two resolutions, 
and capturing the unique elementary topology change in this class of Type IIB 
geometries.  
\fig{The transition geometry can be resolved in two ways, related to each 
other by the particle-hole duality in the fermion 
picture.}{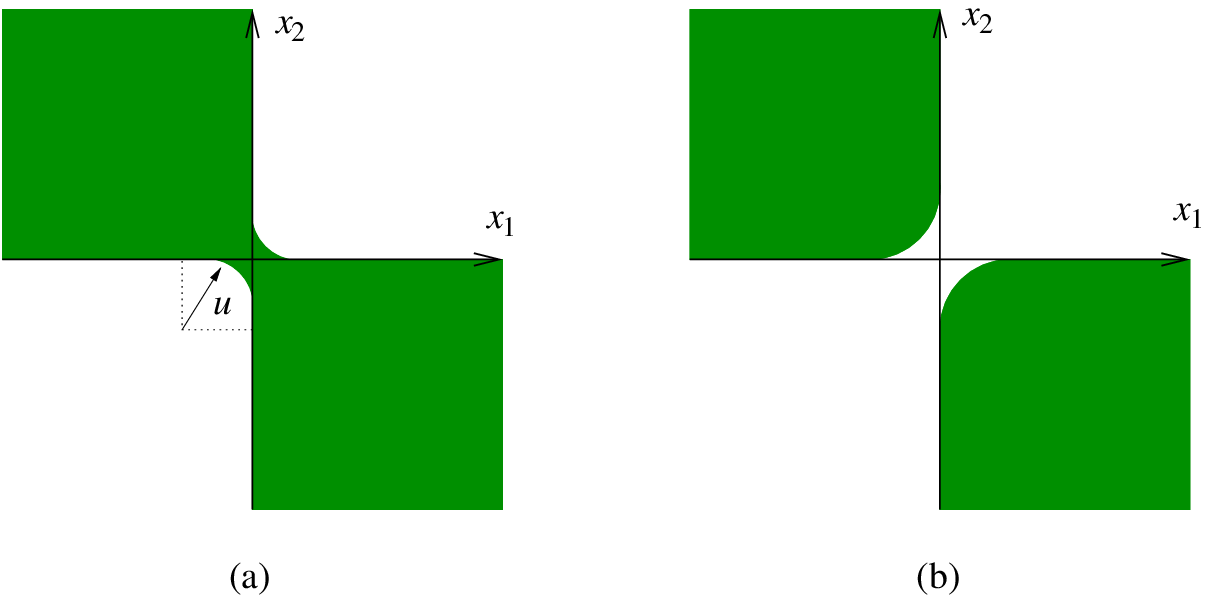}{5truein} 

\subsec{A one-parameter family of resolved geometries}

First we will consider a typical resolved geometry, for which the Fermi 
surface is affected only in some compact region of size $u$ around 
the origin.  For concreteness, we pick the distribution of fermions whose 
Fermi surface is given by $(x_1-u)^2+(x_2-u)^2=u^2$ for $0\leq x_1,x_2\leq u$, 
$(x_1+u)^2+(x_2+u)^2=u^2$ for $-u\leq x_1,x_2\leq 0$, and by $x_1x_2=0$ 
otherwise (Fig.~5(a)).   The positive parameter $u>0$ is a modulus of this 
family of solutions.  

The corresponding function $z$ seeded by this distribution of fermions can 
be analytically evaluated, leading to
\eqn\eezeerescirc{\eqalign{z(x_1,x_2,y)={}\frac{}{2\pi}&\!\frac{x_1}{
\sqrt{x_1^2+y^2}}\arctan\left(\frac{x_2-u}{\sqrt{x_1^2+y^2}}\right)+
\frac{x_2}{2\pi\sqrt{x_2^2+y^2}}
\arctan\left(\frac{x_1-u}{\sqrt{x_2^2+y^2}}\right)\cr
{}+\frac{1}{4}+\frac{}{2\pi}&\!\frac{x_1}{\sqrt{x_1^2+y^2}}
\arctan\left(\frac{x_2+u}{\sqrt{x_1^2+y^2}}\right)+
\frac{x_2}{2\pi\sqrt{x_2^2+y^2}}
\arctan\left(\frac{x_1+u}{\sqrt{x_2^2+y^2}}\right)\cr
{}-\frac{H_-}{2\pi\sqrt{H_-^2+4u^2y^2}}&
\left\{\arctan\left(\frac{\sqrt{H_-^2+4u^2y^2}}{x_1^2+x_2^2+y^2-u^2}\right)
+\frac{\pi}{2}\left[1-\sign(x_1^2+x_2^2+y^2-u^2)\right]\right\}\cr
{}-\frac{H_+}{2\pi\sqrt{H_+^2+4u^2y^2}}&
\left\{\arctan\left(\frac{\sqrt{H_+^2+4u^2y^2}}{x_1^2+x_2^2+y^2-u^2}\right)
+\frac{\pi}{2}\left[1-\sign(x_1^2+x_2^2+y^2-u^2)\right]\right\}\cr}}
where we have introduced a shorthand notation
\eqn\eeshorth{H_{\pm}=(x_1\pm u)^2+(x_2\pm u)^2+y^2-u^2.}

The family that corresponds to Fig.~5(b) can be obtained from \eezeerescirc\ 
by rotating the phase space by $\pi/2$ and invoking the particle-hole 
duality.  We will parametrize this other branch of resolved geometries 
by negative values of $u$, with $-u$ being the size of the resolved region 
in Fig.~5(b).  Interpolating between positive and negative $u$ takes the 
geometry through the topology changing transition.  

The analytic form of this class of half-BPS solutions in the LLM coordinates 
is admittedly not too illuminating, besides illustrating that (a) a family 
can be analytically constructed, and (b) that the generic solution in the 
LLM class is fairly complicated (at least in the LLM coordinates).  

Having resolved the singular geometry in two different ways, we can now 
analyze in more detail the change of topology that this entails as $u$ is 
varied.  At $u\neq 0$, the geometry is non-singular.  As we approach $u=0$,  
the topology of the spatial sections changes, from $R^5\times S^4$ at $u<0$, 
via the cone over $S^4\times S^4$ at $u=0$, to $S^4\times R^5$ at $u>0$.  
During this process of topology change, an $S^4$ cycle in $R^5\times S^4$ 
shrinks to a point (say) at the origin of $R^5$, thus forming a singular cone 
over $S^4\times S^4$ at $u>0$.  Then the singular point blows up into 
another $S^4$, turning the topology of the solution into the nonsingular 
$S^4\times R^5$.  In the process, the contractible and non-contractible 
factors of the geometry exchanged their roles.  This mechanism of topology 
changing transition is notably reminiscent of the behavior near the conifold 
singularities of Calabi-Yau manifolds.  
  
\fig{The resolution that corresponds to the Fermi surface of non-critical 
$\hat c=1$ Type 0B string theory.}{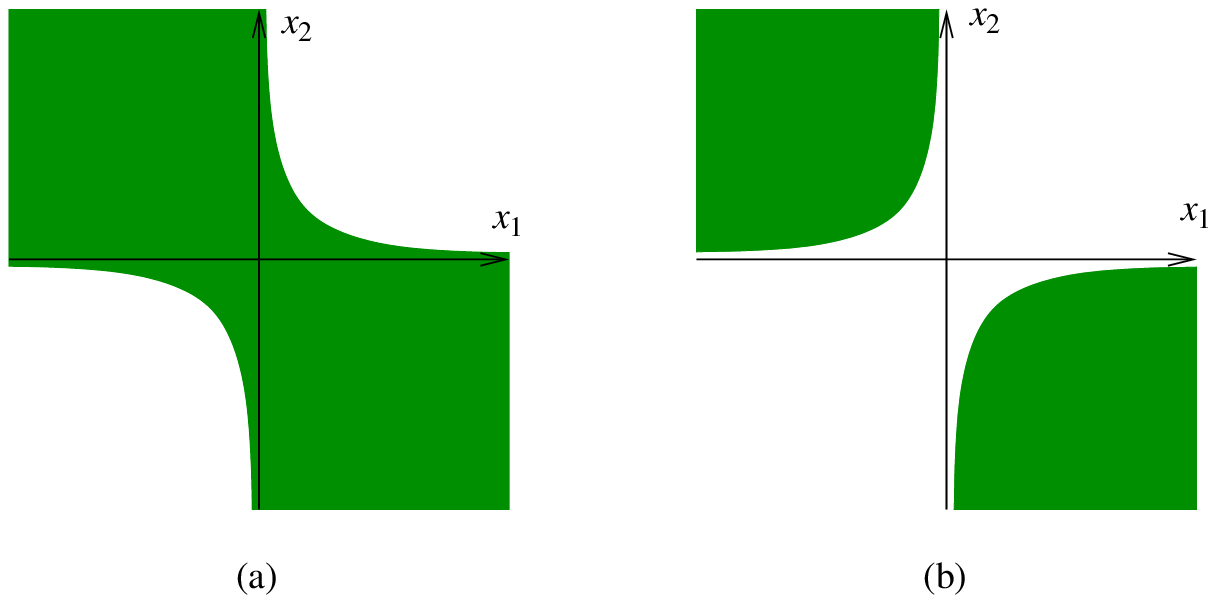}{5truein} 

\subsec{Resolutions and two-dimensional Type 0B strings}

The particular details of how the Fermi surface is separated into two 
disconnected smooth components are immaterial: any such resolved distribution 
of fermions will give rise to a non-singular Type IIB geometry whose topology 
is as described above.  Two resolutions that differ from each other just by 
a ripple on the Fermi surface are related to each other by excitations of the 
supergravity modes on this background.  

Consider a distribution of fermions such that the Fermi surface is given by
\eqn\eeboundmu{x_1x_2=2\mu,}
where positive and negative values of $\mu$ correspond to Fig.~6(a) and 6(b) 
respectively.  When $\mu=0$ we recover our original geometry, while for 
$\mu\neq 0$ this represents another one-parameter family of smooth solutions 
undergoing the topology change as $\mu\rightarrow 0$.  This family is 
topologically equivalent to the one constructed in the previous subsection, 
since their corresponding fermion distributions differ only by a wave on the 
Fermi surface.  

After a simple redefinition of coordinates, 
\eqn\eepqphase{p=\frac{1}{\sqrt{2}}(x_1+x_2),\qquad q=\frac{1}{\sqrt{2}}
(x_1-x_2).}
the Fermi surface \eeboundmu\ can be written as 
\eqn\eefermicone{p^2-q^2=\mu.}
This is the Fermi surface of the much studied two-dimensional non-critical 
string theory (see \crev\ for reviews), in the canonical phase space 
coordinates.  In fact, since both sides of the Fermi sea are filled, this is 
more precisely interpreted as the Fermi sea of the ground state of the 
$\hat c=1$ Type 0B string theory, as described by the matrix model with the 
inverted harmonic oscillator potential \chat.  

Noncritical string theory has appeared in the description of critical string 
backgrounds near singularities before (see, e.g.,~\refs{\vafaetc -\minaetal}). 
In our case, it is the full Type 0B theory in two dimensions with 
uncompactified, Lorentz-signature time that makes its appearance, not just 
the topological or Euclidean version of this noncritical string theory.  

This appearance of the Fermi surface of the inverted harmonic oscillator in 
the description of the half-BPS Type IIB geometry near its topological 
transition can be naturally interpreted on the CFT side of the AdS/CFT 
correspondence as follows.  In CFT, the half-BPS operators with $\Delta=J$ are 
described by the gauged matrix model \refs{\antal-\calda} with action 
\eqn\eemmaction{S_0=\frac{1}{2}\int dt\,\Tr\left\{(D_t\Phi)^2-\Phi^2\right\},} 
with $\Phi$ a hermitian matrix.%
\foot{For an explanation of how the complex matrix model of the zero modes of 
$Z=\phi_1+i\phi_2$ in CFT reduces in the half-BPS sector with $\Delta=J$ to 
a hermitian matrix model, see \antal .}
The natural ground state of the fermions in this matrix model is the circular 
droplet that corresponds to $AdS_5\times S^5$ on the Type IIB side.  Composite 
operators $\Pi_i(\Tr\,\Phi^{n_i})^{r_i}$ are the natural observables of this 
matrix model.  These observables naturally lead to a multi-parameter family of 
deformations of the matrix model,
\eqn\eemmactdef{S(g_{n_ir_i})=S_0+\int dt\sum g_{n_ir_i}
\Pi_i(\Tr\,\,\Phi^{n_i})^{r_i},} 
parametrized by arbitrary coupling constants $g_{n_ir_i}$.  In the 
construction of the corresponding half-BPS Type IIB geometries \llm , 
the matrix model \eemmaction\ with the harmonic oscillator potential does 
not play any privileged role.  All distributions of fermions on the $x_1,x_2$ 
appear democratically.  Some distributions of fermions will be small 
perturbations of the $AdS_5\times S^5$ ground state; those are naturally 
interpreted as finite-energy excitations of the harmonic oscillator theory 
\eemmaction .  Other distributions are far from the $AdS_5\times S^5$ ground 
state.  Our geometry near the topological transition is an example of that.  
Instead of interpreting such solutions as highly excited state of 
$AdS_5\times S^5$, it is more natural to think of them as ground states of 
a deformed matrix model \eemmactdef .  In particular, our topology changing 
solution, whose Fermi surface is given by \eefermicone , is naturally 
interpreted as the ground state of the matrix model with the inverted harmonic 
oscillator potential, 
\eqn\eemmaction{S=\frac{1}{2}\int dt\,\Tr\left\{(D_t\Phi)^2+\Phi^2\right\}.} 
This matrix model captures more naturally the physics of half-BPS 
perturbations near the ground state of the geometry near topological 
transition.  

\newsec{Finite $N$: Topology Change Inside $AdS_5\times S^5$}

Until now we focused on the universal features of the unique topology changing 
process possible in the LLM geometries, by zooming in on a small neighborhood 
of the singularity.  Now we wish to study the topology changing process 
embedded into a larger environment, for example, in asymptotic 
$AdS_5\times S^5$.  

\subsec{Topology Change in Asymptotic $AdS_5\times S^5$}

Embedding our topology change into $AdS_5\times S^5$ requires restoring finite 
$N$, and considering a distribution of fermions confined into a compact region 
in the $x_1,x_2$ plane.  One particularly interesting distribution of 
fermions with such features is schematically illustrated in Fig.~7.  In the 
$p,q$ coordinates of \eepqphase , we consider a Fermi surface given by
\eqn\eefermifour{\frac{1}{2}p^2-\frac{1}{2}q^2+\frac{g}{4}q^4=\mu}
\fig{The topology changing transition inside $AdS_5\times S^5$. Here (a) 
corresponds to the Fermi surface \eefermifour\ with $\mu<0$, and (b) describes 
$\mu>0$.}{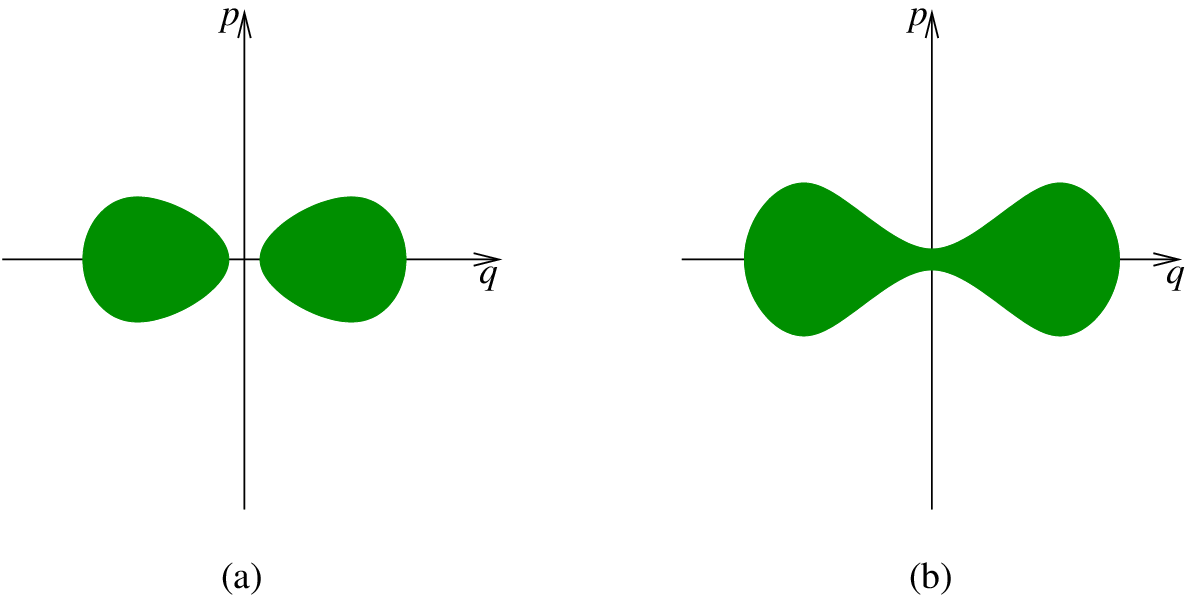}{5truein} 
Here $\mu$ plays the usual role of a chemical potential for the fermions, 
with its value determined in terms of $N$ and the quartic coupling constant 
$g$ by requiring 
\eqn\eemuchem{N=\int_\CD\frac{dp\,dq}{4\pi^2\ell_p^4},}
where the integral is performed over the filled droplet(s) $\CD$ in the $p,q$ 
space.  

The $\mu$-dependent family of geometries that result from the distribution of 
fermions with Fermi surface \eefermifour \ describes the macroscopic topology 
change of half-BPS geometries asymptotic to $AdS_5\times S^5$.  For $\mu>0$, 
we have one connected droplet of fermions, and the topology of the solution 
is still that of $AdS_5\times S^5$.  However, geometrically, near the origin 
of $AdS_5$, the metric of the $S^5$ factor has been smoothly distorted by 
fluxes so that its equatorial $S^4$ is shrinking to zero size.  At the point 
of the topological transition at $\mu=0$, this $S^4$ has shrunk to a point 
precisely at the origin in $AdS_5$, pinching the original $S^5$ there into two 
topologically distinct $S^5$'s. When we continue to $\mu<0$, these two $S^5$'s 
will separate, in a process that can be described as the conical singularity 
of the solution at $\mu=0$ blowing up into a noncontractible $S^4$.  
This noncontractible $S^4$ can be effectively thought of as the origin of 
$AdS_5$ blown up into an $S^3$, with the remaining angular coordinate on $S^4$ 
parametrizing the distance between the two $S^5$s.  Zooming in on the 
singularity of this solution again reproduces our statement that the spatial 
slice of the singular solution at infinite $N$ is a cone over 
$S^4\times S^4$.  This topology change can be viewed as an 
$AdS_5\times S^5$ version of anti-de~Sitter fragmentation \frag .

In the matrix model language, the geometry near its topological transition can 
be interpreted as a highly excited state in the original matrix model 
\eemmaction , or as the ground state of a deformed matrix model \eemmactdef , 
\eqn\eemminads{S(g)=\frac{1}{2}\int dt\,\Tr\left\{(D_t\Phi)^2+\Phi^2 
-\frac{g}{2}\Phi^4\right\}.} 
related to the original harmonic oscillator model \eemmaction\ by adding two 
single-trace observables $\Tr(\Phi^2)$ and $\Tr(\Phi^4)$ to the action.   

\subsec{The Penrose limit as the double scaling limit of noncritical Type 
0B theory}

It is interesting to take another look at the limit of the $AdS_5\times S^5$ 
geometry, near its topological transition, which reproduces the universal 
geometry studied in Sections~2-4.  From the point of view of the asymptotic 
$AdS_5\times S^5$ geometry, this involves zooming in on the vicinity of the 
singularity, and simultaneously scaling $\mu$ such that the effective distance 
between the two components of the Fermi surface stays finite.  This can also 
be seen as an $N\rightarrow\infty$ limit, in which we recover the distribution 
of fermions of Fig.~6.  

It turns out that this near-singularity limit has a very natural 
interpretation, both in supergravity and in CFT. 

On the supergravity side, this limit simply corresponds to the Penrose limit.  
Indeed, notice first that at $\mu=0$, the singularity itself is a null 
line in the full ten-dimensional geometry.  Moreover, for $\mu\neq 0$, one 
can mark a point $P$ on the boundary between the filled and empty regions that 
is near the origin in the $p,q$ coordinates at $y=0$.  In the full geometry, 
$P$ corresponds to a null geodesic.  The Penrose limit towards this geodesic 
corresponds to shifting the origin of the $x_1,x_2$ coordinate system to $P$,  
rescaling uniformly 
\eqn\eepenrose{(x_1,x_2,y)\rightarrow\frac{1}{\lambda}(x_1,x_2,y)}
while keeping $t$ constant, and then setting $\lambda=R_{AdS}^2\sim N^{1/2}$ 
and sending it to infinity.  This alone would indeed be the conventional 
Penrose limit in the LLM coordinates, and would result in the maximally 
symmetric pp-wave.  Here, however, we are interested in a combined limit, 
which also rescales 
\eqn\eemuresc{\mu\rightarrow\frac{\mu}{\lambda^2},}
ensuring that the rescaled distance between the two components of the Fermi 
surface stays finite as we take the Penrose limit $N\rightarrow\infty$.  
In this way, we stay near the (resolved) singularity during the Penrose 
limit.  The distribution of fermions that corresponds to this Penrose limit 
of the original $AdS_5\times S^5$ near its topological transition is that of 
Fig.~6. 

On the CFT side, the double scaling limit \eepenrose\ and \eemuresc\ 
is precisely the conventional double-scaling limit of the matrix model 
\eemminads (see e.g.~\crev\ for a review of $c=1$ strings and the double 
scaling limit).  In this limit, the $\Phi^4$ term in the potential of the 
matrix model is effectively scaled away, and we are left with the theory of 
double-scaled fermions in the inverted harmonic oscillator potential -- the 
theory that defines the noncritical $\hat c=1$ Type 0B string theory.  After 
the double scaling limit is taken, the effective string coupling 
$g_{s,0{\rm B}}$ of the two-dimensional Type 0B theory is set by the 
distance between the Fermi surface from the top of the inverted harmonic 
oscillator potential,
\eqn\eegstreff{g_{s,0{\rm B}}\sim\frac{1}{\mu}.}
It is pleasing to see that the two natural limits -- the Penrose limit in 
supergravity and the double-scaling limit in the matrix model -- are related 
to each other in this simple way.  

Using this relation between the Penrose limit and the double-scaling 
limit, we expect the scattering amplitudes of the massless tachyon and RR 
scalar modes of Type 0B string theory in two dimensions to encode the 
amplitudes for the propagation of half-BPS supergravity modes through our 
Type IIB solution near its topological transition, i.e., the natural 
``background S-matrix'' observables proposed at the end of Section~3.1.

\subsec{Energetics of Topology Change in $AdS_5\times S^5$}

The geometry represented by the Fermi distribution of \eefermifour\  (or 
Fig.~7) is asymptotically $AdS_5\times S_5$.  However, it does not represent 
a small perturbation of the $AdS_5\times S^5$ vacuum.  Half-BPS excitations 
of the vacuum with relatively low energies would be, for example, supergravity 
modes (with $\Delta\ll N$) or D-branes known as giant gravitons (with $\Delta 
\sim N$).  Our geometry is instead a smooth geometry with fluxes, and cannot 
be usefully  approximated as a distribution of some small number of branes 
inside $AdS_5\times S^5$.  In fact, using the formula for $\Delta$ \llm 
\eqn\eedeltacom{\Delta=J=\frac{1}{2}\int_\CD\frac{dp\,dq}{8\pi^3\ell_p^8}(p^2+
q^2) -\frac{1}{2}\left(\int_\CD\frac{dp\,dq}{4\pi^2\ell_p^4}\right)^2}
one can easily show that our geometry is an excitation of $AdS_5\times S^5$ 
with $\Delta\sim N^2$, and therefore represents a qualitatively new effect 
in $AdS_5\times S^5$.  

Similarly, the Penrose limit of our solution which was the subject of Sections 
2-4 is not a low-energy excitation of the pp-wave, i.e., it cannot be obtained 
from the latter by adding a finite number of supergravity modes or giant 
gravitons.  

Having argued that the generic topology-changing solution in asymptotic 
$AdS_5\times S^5$ has $\Delta\sim N^2$, one can ask: what is the 
{\it lowest\/} value of $\Delta$ needed for a macroscopic topology change of 
$AdS_5\times S^5$, at least within this class of half-BPS geometries?  
As we will now show, the answer turns out to be of order $\Delta\sim N^{3/2}$. 
This energy is parametrically lower than the generic value $\Delta\sim N^2$ 
expexted from a generic smooth LLM geometry, but still higher than $\sim N$, 
which would result from a small collection of branes/giant gravitons in 
$AdS_5\times S^5$.  

In the semiclassical fermion picture, this lowest-$\Delta$ geometry that 
changes the topology  results from starting with the circular Fermi sea of 
radius $\sim N$ that describes the vacuum $AdS_5\times S^5$, and removing 
$\sim\sqrt{N}$ fermions from it along (say) the $x_2=0$ line, thus creating 
a passage in the Fermi sea of the characteristic width $\sim\sqrt{\hbar}$.  
Using \eedeltacom\ to evaluate $\Delta$ for this geometry with the strip 
removed, we obtain $\Delta\sim N^{3/2}$.%
\foot{If one were to insist on preserving the overall value of $N$, the 
removed fermions would then have to be spread on top of the Fermi surface.  
It is easy to check that this also causes an effect of order $N^{3/2}$.}
In the conventional Type IIB language, this creation of $\sim\sqrt{N}$ aligned 
holes in the Fermi sea corresponds to placing $\sim\sqrt{N}$ nested giant 
gravitons on $S^5$.  

\newsec{Other Half-BPS Solutions with Scaling}

Until now we focused on the irreducible topology changing process, in which 
just two components of the Fermi surface meet in a point.  As we argued, any 
other more complicated localized singularity in this class of half-BPS 
geometries can be smoothly decomposed into a sequence of some number of 
such irreducible transitions.  In the process, the topology of the spacetime 
changes, at each step by the simple transition described above.  Clearly, 
applying this process many times can lead to fairly complicated topologies, 
and it would be nice to have a simple tool for analyzing this.    

In this section, we present such a tool, in the form of a simple description 
of all possible localized singularities, irreducible or not, in this class of 
half-BPS solutions in Type IIB theory.  The crucial observation is that the 
universal features of a localized singularity are captured in the generalized 
Penrose limit, in which we take the $N\rightarrow\infty$ limit and/or zoom in 
on a small region near the singularity.  In that limit, the distribution of 
fermions that corresponds to the singular geometry will be invariant under 
uniform rescalings of the $x_1,x_2$ plane, and consequently, the full solution 
has to share the same scaling properties \eescaling\ -- \eemetscaling\ 
discussed in Section~2.3.

The corresponding scale-invariant distributions of fermions can be easily 
classified: they correspond to the Fermi surface consisting of $2k$ 
straight lines meeting at the origin, with $k=1,2,\ldots$, as in Fig.~8.  
\fig{The fermion distribution that corresponds to a $2k-1$-parameter family 
of geometries with scaling.  This family is parametrized by moduli $\alpha_a$, 
$a=1,\ldots, 2k-1$.}{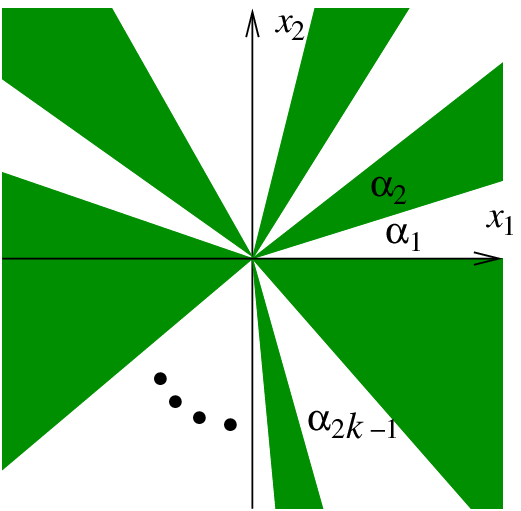}{2truein}
The $2k$ angles $\alpha_a$ between two consecutive (i.e., $a$-th and $a+1$-st) 
lines, subject to the constraint that $\sum_{a=1}^{2k}\alpha_a=2\pi$, 
represent the moduli of this $2k-1$-parameter family of solutions.  The Type 
IIB geometries of this family can also be easily constructed, by first solving 
for $z$ with the boundary conditions set by the fermion distribution.  
Each individual $a$-th line, with $a=1,\ldots 2k$, contributes
\eqn\eeindivline{z_a=\frac{v_a}{2\pi\sqrt{v_a^2+y^2}}\arctan\left({
\frac{u_a}{v_a^2+y^2}}\right)+\frac{v_a}{4\sqrt{v_a^2+y^2}},}
where
\eqn\eeuava{\eqalign{u_a&=x_1\cos\sum_{b=1}^{a-1}\alpha_b+x_2\sin
\sum_{b=1}^{a-1}\alpha_b,\cr
v_a&=x_2\cos\sum_{b=1}^{a-1}\alpha_b-x_1\sin\sum_{b=1}^{a-1}\alpha_b,\cr}}
The total function $z$ for this solution is then
\eqn\eetotalza{z=-\sum_{a=1}^{2k}(-1)^az_a.}
The full metric and fluxes of this Type IIB solution with a null singularity 
is then determined in terms of $z$ using the rules summarized in Section~2.1.  

The only non-singular geometry on this list is that of $k=1$, with two lines 
meeting at the origin at a straight angle, the resulting geometry being the 
non-singular maximally supersymmetric pp-wave.  When the angle between the two 
lines is $\pi+\alpha$ with $\alpha$ small, the pp-wave solution will develop a 
mild singularity due to the small localized disturbance of the Fermi sea at 
the origin.  This geometry again has two asymptotic regimes where it is 
asymptotic to the pp-wave.  Introducing observers and coordinates similar 
to those of Section~3, one can study this perturbed pp-wave as a 
time-dependent solution, asymptotic to the maximally supersymmetric pp-wave 
in the far past and the far future.  It might be interesting to further study 
the physics of this perturbation of the pp-wave.  

Another interesting geometry in the class studied in this section is the 
case with $k=2$, with the four angles given in terms of a small angle $\alpha$ 
by $\alpha_1=\alpha_3=\pi-\alpha$, $\alpha_2=\alpha_4=\alpha$, $\alpha\ll 1$.  
This corresponds to fermions filling the entire plane, with two narrow edges 
cut out from the Fermi sea in two opposite directions.  This geometry is 
closer to being the (Penrose limit of) the configuration that changes the 
macroscopic topology of $AdS_5\times S^5$ at the lowest energy 
$\Delta\sim N^{3/2}$, as discussed in Section~5.3.  

\newsec{Conclusions}  

In this paper we have initiated the study of spacetime topology 
change in LLM geometries.  We have pointed out that all localized topology 
changing transitions are reducible to a unique transition, and presented the 
supersymmetric Type IIB solution that describes the universal features of 
this unique transition.  

We have also shown that this solution can be obtained from a class of topology 
changing solutions in asymptotic $AdS_5\times S^5$ as their Penrose limit.  
On the CFT side, this Penrose limit nicely corresponds to the double scaling 
limit of the matrix model in an inverted harmonic oscillator potential, 
which is known to describe noncritical Type 0B string theory in two 
dimensions.  

The spatial section of the solution at topological transition is a cone over 
$S^4\times S^4$.  The tip of the cone is a supersymmetric, null, point-like 
singularity.  The two ways of resolving the singularity within the half-BPS 
family closely resemble the conifold transition in Calabi-Yau manifolds.  

One fascinating question to ask is whether the singularity can be 
thermalized.  Since this singularity has emerged in the highly controlled 
class of LLM geometries with sixteen supersymmetries, and the free 
Fermi theory provides a prescription for the quantization of such geometries, 
it seems very likely that this singularity should be on the list of ``good'' 
spacetime singularities that string theory knows how to make sense of.  One 
general feature expected of ``good'' singularities in string theory is that 
they can be thermalized \gubser\ -- when we move the solution slightly away 
from BPS, the singularity turns into a thermal object, typically carrying 
macroscopic entropy.  In this sense, this thermalization of the singularity 
is expected to lead to the formation of a macroscopic black-hole horizon.  
It would be very interesting to study near-BPS versions of our singular 
geometry, and see whether a finite-area horizon is grown.  A natural class of 
near-BPS states would correspond to the complex matrix model of \antal\ with 
both types of oscillators possibly excited, so that the resulting 
configuration no longer satisfies the BPS condition $\Delta=J$ but the matrix 
model framework still applies. 

\bigskip
\bigskip
\noindent{\bf Acknowledgements}
\medskip
We wish to thank Mina~Aganagic, David~Berenstein, Antal~Jevicki and 
Aleksey~Mints for useful discussions and comments.  One of us (PGS) also 
benefitted from helpful discussions with Nadir~Jeevanjee.  This material is 
based upon work supported in part by NSF grant PHY-0244900, by the Berkeley 
Center for Theoretical Physics, and by DOE grant DE-AC03-76SF00098.  Any 
opinions, findings, and conclusions or recommendations expressed in this 
material are those of the author(s) and do not necessarily reflect the views 
of the National Science Foundation.  
%%%%%%%%%%%%%%%%%%%%%%%%%%%%%%%%%%%%%%%%%%%%%%%%%%%%%%%%%%%%%%%%%%%%%%%%%%%%%
\listrefs
\end